# Silicene-based DNA Nucleobase Sensing

Hatef Sadeghi, S. Bailey and Colin J. Lambert[*]
*Lancaster Quantum Technology Center, Physics Department, Lancaster University, LA1 4YB Lancaster, UK*

We propose a DNA sequencing scheme based on silicene nanopores. Using first principles theory, we compute the electrical properties of such pores in the absence and presence of nucleobases. Within a two-terminal geometry, we analyze the current-voltage relation in the presence of nucleobases with various orientations. We demonstrate that when nucleobases pass through a pore, even after sampling over many orientations, changes in the electrical properties of the ribbon can be used to discriminate between bases.

DNA sequencing (sensing the order of bases in a DNA strand) is an essential step toward personalized medicine for improving human health[1]. Despite recent developments, conventional DNA sequencing methods are still expensive and time consuming[2]. Therefore the challenge of developing accurate, fast and inexpensive, fourth-generation DNA sequencing alternatives has attracted huge scientific interest[3]. All molecular based biosensors rely on a molecular recognition layer and a signal transducer, which converts specific recognition events into optical, mechanical, electrochemical or electrical signals[4]. Of these, electrical transduction is potentially the fastest and least expensive, because it is compatible with nanoelectronics integration technologies. However, attempts to realize such sensors based on silicon platforms, silicon nanowires or graphene[5,6] have not yet delivered the required level of selectivity. In this paper, we examine the potential of the recently-synthesized two-dimensional material silicene as a platform for DNA sequencing and demonstrate that the unique electrical properties of nanoporous silicene allow direct electrical transduction and selective sensing of nucleobases.

Silicene (fig. 1a) is a recently-observed one-atom-thick crystalline form of silicon with $sp^2$ bonded atoms arranged in a slightly buckled honeycomb lattice structure[7,8-10]. The synthesis of silicene nanoribbons has been demonstrated on silver (111)[8,11,12,13], gold (110)[14], iridium (111)[15] and the zirconium diboride (0001)[16] substrates. In contrast with graphene, the buckling of silicene[17] can open up an energy gap at the Fermi energy $E_F$ of between 300meV[13] and 800meV[10], which can be controlled by an external perpendicular electric field[9]. Silicene[12] and silicene nanoribbon edges are also chemically stable to $O_2$ exposure[18].

Given the compatibility of silicene with existing semiconductor techniques, it is natural to ask if this material can form a platform for DNA sequencing and therefore in what follows we examine the potential of nanoporous silicene nanoribbons for direct electrical sensing of nucleobases. Current technology allows the drilling of nanopores with different diameters down to a few angstroms in graphene, $Al_2O_3$ and $TiO_2$ based membranes[19]. Three types of nanopores have been presented in the literature for DNA sensing[20]. Currently-available solid-state nanopore-based strategies rely on reading the variation of an ionic current through a surrounding fluid due to the translocation of DNA strand through a pore in a solid state membrane[21]. However, ionic current leakage in the thin membranes, poor signal to noise and difficulties in controlling the speed of translocation through the pore have so far limited the development of this technique[22]. In a second approach, biological nanopores (MspA and α-hemolysin) have been employed as recognition sites inside the pore[23]. This method overcomes key technical problems required for real-time, high-resolution nucleotide monophosphate detection[24], but several outstanding issues need to be addressed, including the sensitivity of biological nanopores to experimental conditions, the difficulty in integrating biological systems into large-scale arrays, the very small (~pA) ionic currents and the mechanical instability of the lipid bilayer that supports the nanopore[6,25]. As a third approach to nanopore-based sequencing, changes in the electrical conductance of single-layer graphene have been used for DNA sensing[26,27]. In general, direct electrical conductance measurement is more attractive than blockade ionic current measurement, since the response of the former is much faster and the signal to noise is higher. However, monolayer graphene does not show sufficient selectivity[27].

---

[*] Author to whom correspondence should be addressed. Email: c.lambert@lancaster.ac.uk

Here we propose silicene nanopores for DNA sequencing and demonstrate that the electrical conductance of silicene nanoribbons containing a nanopore is selectively sensitive to the translocation of DNA bases through the pore. An example of a silicene nanoribbon containing a nanopore is shown in figure 1b. The electrical conductance *G* of this nanopore-containing silicene ribbon is computed using a first-principles quantum transport method, implemented using the well-established codes SIESTA[28] and SMEAGOL[29]. This involves computing the transmission coefficient *T(E)* for electrons of energy *E* passing from a source on the left to a drain on the right through the structure shown in figure 1b.

To find the optimized geometry and ground state Hamiltonian of each system, we employed the SIESTA[28] implementation of density functional theory (DFT) within the generalized gradient approximation (GGA) correlation functional with the Perdew-Burke-Ernzerhof parameterization (PBE). Results were found to converge with a double zeta polarized basis set, a plane wave cut-off energy of 250 Ry and a maximum force tolerance of 20 meV/Ang. k-point sampling of the Brillion zone was performed by 1×1×20 Monkhorst–Pack grid. Using the Hamiltonian obtained from DFT, the Green's function of the open system (connected to silicene leads) is constructed and the transport calculation performed using the SMEAGOL implementation of non-equilibrium Green's functions[29].

To use the non-equilibrium Green's function formalism, the Hamiltonian of the whole pore-containing ribbon is needed, both in the presence and absence of nucleotides. The converged profile of charge via the self-consistent DFT loop for the density matrix implemented by SIESTA is used to obtain this Hamiltonian. Employing the SMEAGOL method [29], the transmission coefficient between two lead in two terminal system is then given by: $T(E) = Trace\{\Gamma_R(E)G^R(E)\Gamma_L(E)G^{R\dagger}(E)\}$ where $\Gamma_{L,R}(E) = i(\sum_{L,R}(E) - \sum_{L,R}^{\dagger}(E))$ are the level broadening due to the coupling between left and right electrodes and the scatter, $\sum_{L,R}(E)$ are the retarded self-energies of the left and right leads and $G^R = (ES - H - \sum_L - \sum_R)^{-1}$ is retarded Green's function, where *H* and *S* are Hamiltonian (obtained from the DFT self-consistent loop implemented by SIESTA) and overlap matrices, respectively.

For a perfect nanoribbon (ie in the absence of a nanopore), figure 1c shows the variation of *T(E)* with energy. In this case, the de Broglie waves of electrons travelling from left to right are not scattered and *T(E)* is an integer equal to the number of open scattering channels available to right-moving electrons. The presence of a sharp feature near the (un-gated) Fermi energy (which we define to be $E_F$=0) is a consequence of the unique band structure of silicene nanoribbons and is associated with edge states. In the presence of a nanopore, the resulting *T(E)* is shown in figure 1d. In this case electrons are scattered by the nanopore and *T(E)* is reduced compared to that of the perfect ribbon. Nevertheless the feature near *E*=0 survives.

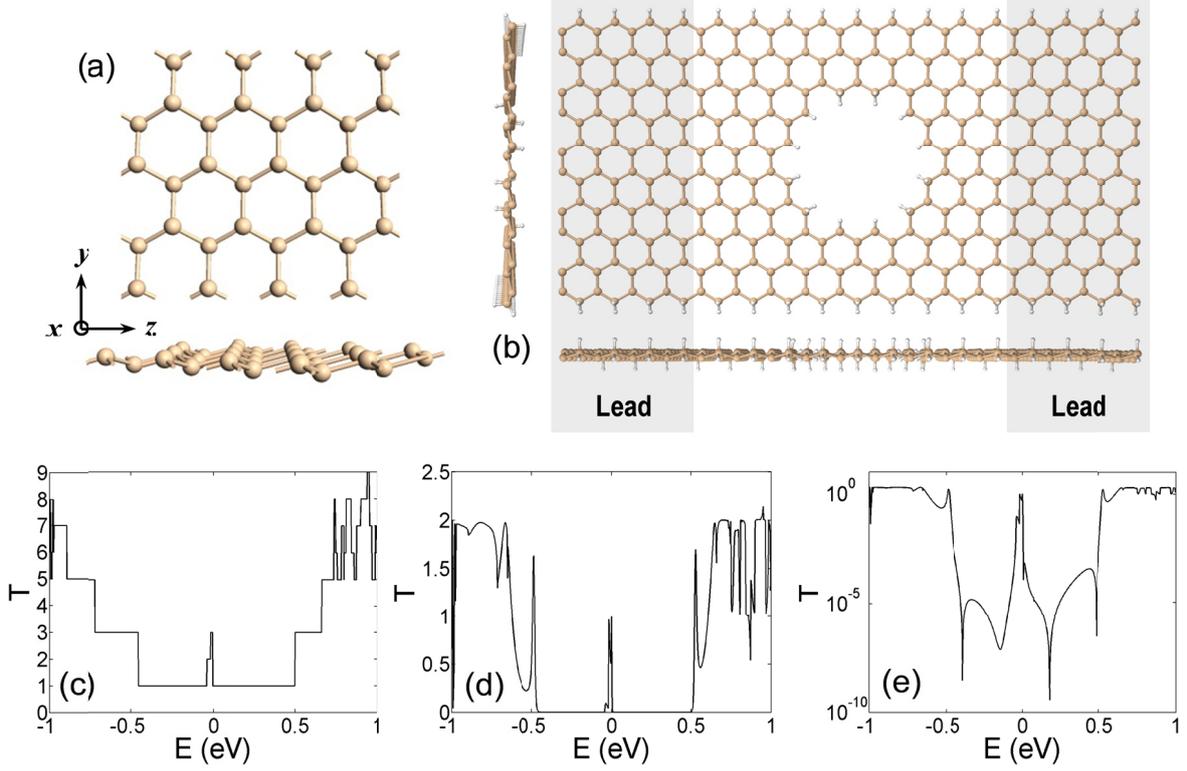

Fig. 1: (a) Silicene molecular structure (b) Molecular structure of monolayer Silicene Nanopore with hydrogen termination in the edges, (c) Transmission coefficient $T_{bare}(E)$ from left lead to the right lead in the absence of a pore (perfect silicene nanoribbon). (d) Transmission coefficient $T_{bare}(E)$ from left lead to the right lead in the presence of a pore. (e) For comparison with figures 3-6, this figure shows a graph of $log10\ T_{bare}(E)$, obtained by plotting figure 1d on a log scale.

In what follows, we compute $T(E)$ in the presence of each of the four bases $X=[A, C, G, T]$. Of course the result depends on the orientation of the base within the pore and therefore for each base $X$, we also consider a number ($m_{max}$) of distinct orientations labeled $m=1, \ldots m_{max}$. The resulting transmission coefficients are denoted $T_{X,m}(E)$. To achieve the required selectivity, an appropriate signal-processing method is required. The most appropriate method will depend on the precise experimental setup, but inevitably will involve interrogating $T_{X,m}(E)$ over a range of energies. An example of such a signal processing method, we examine the following quantity, which can be measured using two-probe geometries, such as that shown in figure 1b:

$$\beta_{X,m}(V) = \log_{10}(I_{X,m}(V)) - \log_{10}(I_{bare}(V)) \tag{1}$$

where $I_{X,m}(V)$ is the current through the device at voltage $V$, in the presence of nucleobase $X$, with orientation $m$, defined by:

$$I_{X,m}(V) = \frac{2e}{h} \int_{E_F - \frac{eV}{2}}^{E_F + \frac{eV}{2}} dE\, T_{X,m}(E) \tag{2}$$

In equation (1), the quantity $I_{bare}(V)$ is the current through the 'bare' device in the absence of any nucleobase. The probability distribution of the set $\{\beta_{X,m}(V)\}$ for a given base X is then defined by

$$P_X(\beta) = \frac{1}{m_{max}(eV_{max} - eV_{min})} \sum_{n=1}^{m_{max}} \int_{eV_{min}}^{eV_{max}} dV\, \delta(\beta - \beta_{X,m}(V)) \tag{3}$$

Alternative discriminators (ie three-terminal device) can also be envisaged, depending on the precise experimental configuration of source, drain and possibly gate electrodes, as discussed in the Supplementary Information[30].

The nanopore of figure 1b has a diameter of 1.7 nm and is created in a zigzag silicene nanoribbon of width 3.2 nm. The edges of the ribbon and the pore are terminated by hydrogen and the structure relaxed to achieve its ground state energy as explained above. We now consider the transmission coefficient of the nanopore upon translocation of nucleotides inside the pore. For each nucleobase, results are presented for $m_{max}$=4 different orientations. Figure 2a shows four orientations of the nucleobase adenine (X=A), inside a silicene pore. The positions and orientations within the pore are obtained by starting from an initial position and orientation and then relaxing the whole structure using the SIESTA implementation of density functional theory. The local geometry of both the surrounding silicene and hydrogen terminations are also relaxed. The resulting structures reveal that all bases are attracted to the surface of the pore, rather than residing near the centre. Once the local energy minima are achieved, the underlying mean field Hamiltonian is used to compute the scattering matrix for de Broglie waves travelling from left to right and from the scattering matrix, the transmission coefficient $T_{Am}(E)$ is obtained. For each of the adenine-containing pores shown in figure 2a, figure 2b shows the corresponding plots of $T_{Am}(E)$. These are used to obtain $I_{Am}(V)$ via equation (2) and are combined with $I_{bare}(V)$ (obtained from $T_{bare}(E)$ of figure 1d or 1e) to yield $\beta_{X,m}(V)$ for X=A.

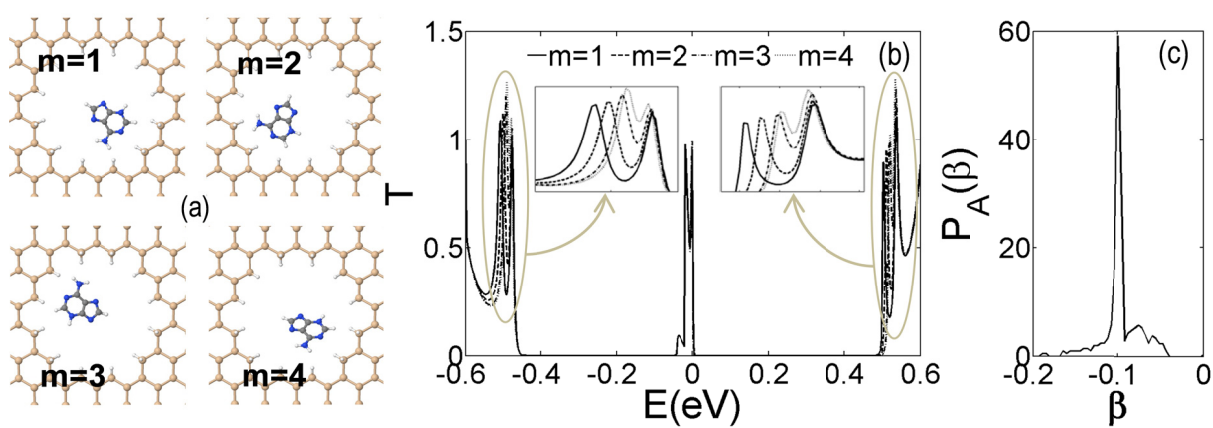

Fig. 2: Fig (a) shows four relaxed geometries and orientations labeled m= 1,2,3,4 of adenine (X=A) within a silicene nanopore. Fig (b and c) show the corresponding plots of $T_{A,m}(E)$ and the probability distribution $P_X(\beta)$. The insets in Fig. 2b show magnifications of the band-edge structure. The maximum voltage employed in the calculation is 0.55 V; meaning that the voltage window is [-0.55 0.55] V.

Figure 2b shows that there are slight differences in the transmissions coefficients for different orientations. When combined together, these lead to the probability distribution $P_A(\beta)$ shown in figure 2c. The changes in transmission coefficients shown in figure 2b (compared with fig 1d) arise from both pi-pi interactions between the adenine and the pore surface states and through electrostatic interactions with the inhomogeneous charge distribution of the nucleobase. These interactions are different for the four bases and ultimately underpin the selectivity demonstrated below. Results for the remaining three bases thymine (X=T), guanine (X=G) and cytosine (X=C) are shown in figures S1, S2 and S3 of the supplementary information[30].

Clearly the transmission coefficients depend on the position and orientation of the nucleobase and therefore the key issue is whether or not this dependency restricts the ability to selectively sense nucleobases within the pore. Figure 3 demonstrates that despite the sensitivity to position and orientation, selective sensing is preserved. For each of the nucleobases, figure 3 shows plots of the quantity $P_x(\beta)$ defined in equation (3). Clearly the presence of well-separated peaks demonstrates that through an appropriate signal processing method, the bases can be selectively detected. The heights and positions of the peaks are different for a given base and either of them could be used to select and recognize the base type. This figure demonstrates the excellent potential of silicene nanopores for DNA sequencing.

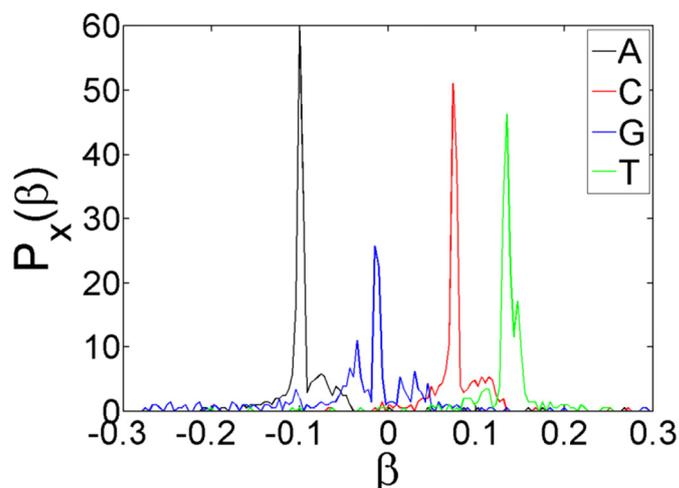

Fig. 3. The probability distribution $P_X(\beta)$ of the set $\{\beta_{X,m}\}$ are shown for a given base X where X = A in black, C in red, G in blue and T in green. The maximum voltage employed in the calculation is 0.55 V; meaning that the voltage window is [-0.55 0.55] V.

In summary, silicene is a material, whose potential applications are only now beginning to be explored. Compared with other two-dimensional materials, it has the immediate advantage of being compatible with existing silicon CMOS technologies. We have performed first principles calculations combined with quantum scattering theory to demonstrate that with appropriate signal processing, silicene-based nanopore sensing offers a potential route to selective sensing of DNA nucleobases. Such a sensing platform is a direct electrical sensor and opens the avenues towards fast, cheap and portable DNA sequencing. In practice there is likely to be variability in pore sizes and shapes and each pore would need to be calibrated prior to use. In this regard, CMOS compatibility is again advantageous, since the potential to create millions of sensors on a single chip, integrated into the necessary control electronics will allow this process can be automated. Furthermore the availability of arrays of nanopores will potentially allow additional refinements in signal processing, leading to further increases in sensitivity and selectivity.

This work is supported by the European Union Marie-Curie Networks NanoCTM and FUNMOLS. Funding is also provided by the UK EPSRC.

# Supplementary Information

# Silicene-based DNA Nucleobase Sensing

Hatef Sadeghi, S. Bailey and Colin J. Lambert[*]

*Lancaster Quantum Technology Center, Physics Department, Lancaster University, LA14YB Lancaster, UK*

Results for transmission coefficients in the presence of X= T, G, C

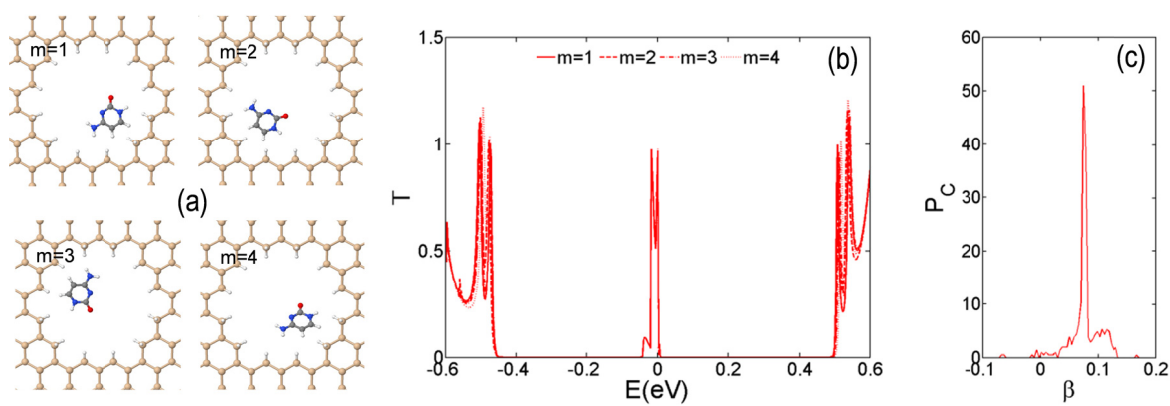

Fig. S1: Fig (a) shows four relaxed geometries and orientations labeled m= 1,2,3,4 of cytosine (X=C) within a silicene nanopore. Fig (b and c) show the corresponding plots of $T_{C,m}(E)$ and the probability distribution $P_C(\beta)$. The maximum voltage employed in the calculation is 0.55 V; meaning that the voltage window is [-0.55 0.55] V.

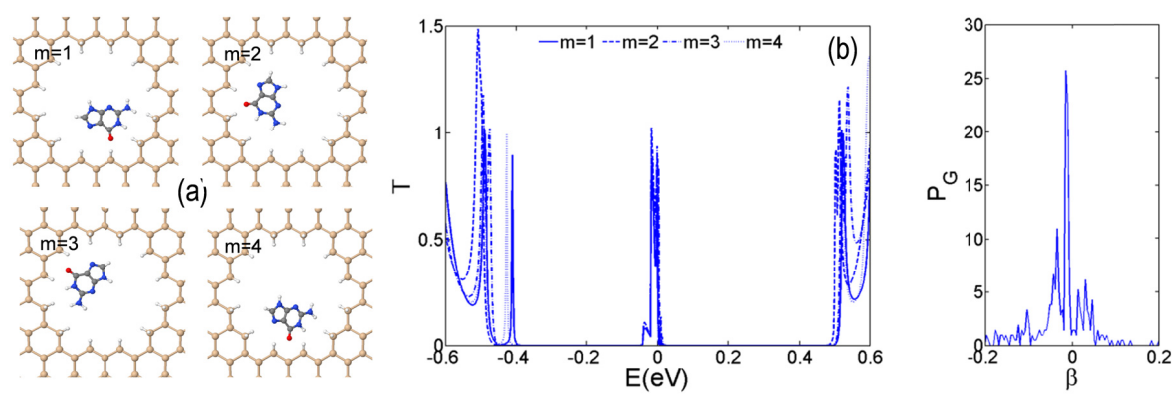

Fig. S2: Fig (a) shows four relaxed geometries and orientations labeled m= 1,2,3,4 of guanine (X=G) within a silicene nanopore. Fig (b and c) show the corresponding plots of $T_{G,m}(E)$ and the probability distribution $P_G(\beta)$. The maximum voltage employed in the calculation is 0.55 V; meaning that the voltage window is [-0.55 0.55] V.

---

[*] Author to whom correspondence should be addressed. Email: c.lambert@lancaster.ac.uk

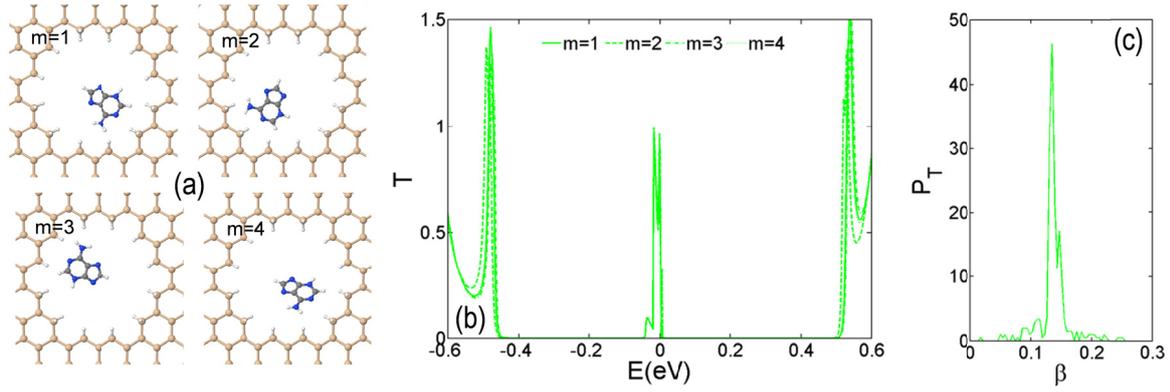

Fig. S3: Fig (a) shows four relaxed geometries and orientations labeled m= 1,2,3,4 of thymine (X=T) within a silicene nanopore. Fig (b and c) show the corresponding plots of $T_{T,m}(E)$ and the probability distribution $P_T(\beta)$. The maximum voltage employed in the calculation is 0.55 V; meaning that the voltage window is [-0.55 0.55] V.

**An alternative signal processing method**

At low source-drain biases, the electrical conductance is given by the Landauer formula $G = G_0 \int_{-\infty}^{+\infty} dE\, T(E)(-\partial f/\partial E)$, where $G_0 = 2e^2/h$ is the conductance quantum and $f(E) = (1 + exp((E - E_F)/k_B T))^{-1}$ is the Fermi function. Provided $T(E)$ is slowly varying on the scale of 0.025eV (ie on the scale of $k_B$ x room temperature) this equation simplifies to $G = G_0\, T(E_F)$. In a real device, $E_F$ can be varied through the application of an external gate potential and therefore $T(E)$ can be sampled over a range of energies. For such a three-terminal device, as an alternative signal processing method, we define the quantity:

$$\alpha_{X,m}(E_F) = \log_{10}(T_{X,m}(E_F)) - \log_{10}(T_{bare}(E_F)) \quad (S1)$$

which is a measure of the differences between $T_{X,m}(E)$ and the transmission of the unoccupied pore in the absence of a base. A plot of the $\log_{10}T_{bare}(E)$ is shown in figure 1e. To differentiate between different bases, we analyze the set of all values of $\alpha_{X,m}(E)$ for energies lying within some convenient range, $E_{min}<E<E_{max}$ and introduce the probability distribution $P_{X,m}(\alpha)$ defined such that $P_{X,m}(\alpha)d\alpha$ is the probability that $\alpha_{X,m}(E)$ lies between $\alpha$ and $\alpha+d\alpha$. Formally this is defined by:

$$P_X(\alpha) = \frac{1}{m_{max}(E_{max} - E_{min})} \sum_{n=1}^{m_{max}} \int_{E_{min}}^{E_{max}} dE\, \delta(\alpha - \alpha_{X,m}(E)) \quad (S2)$$

For each of the nucleobases, figure S4 shows plots of the quantity $P_X(\alpha)$ defined in equation (S2). Well-separated peaks in this figure also show selective sensitivity to the nucleotides regardless of their different orientation.

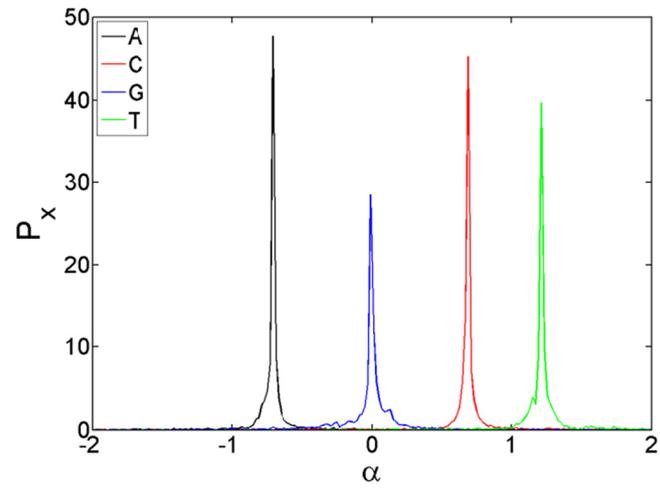

Fig. S4: The probability distribution of the set {α$_{X,m}$(E)} for each nucleobase.